# A variation of the Dragulescu-Yakovenko income model

**José María Sarabia[1], Faustino Prieto, Vanesa Jordá.**

*Department of Economics, University of Cantabria, Avenida de los Castros s/n, 39005 Santander, Spain*

**Abstract**

In the context of the Dragulescu-Yakovenko (2000) model, we show that empirical income distribution with truncated datasets, cannot be properly modeled by the one-parameter exponential distribution. However, a truncated version characterized by an exponential distribution with two parameters gives an accurate fit.

## 1 Introduction

According to Dragulescu and Yakovenko (2000), the fundamental law of equilibrium statistical mechanics is the Boltzmann-Gibbs law, which states that the probability distribution of energy is of the type $P(z) = C\exp(-z/T)$, where T is the temperature, and C is a normalizing constant. Using this result and assuming a closed economic system, these authors demonstrate that, for the countries in the steady state, income follows an exponential distribution.

The present paper shows that empirical individual income distribution cannot be properly modeled by the classical exponential distribution with only one parameter (Cho, 2014; Dragulescu and Yakovenko, 2001a, b) when we have truncated data. This situation appears, for example, with a dataset including only taxpayers. We found that two-parameter exponential distribution can be the proper model instead of one-parameter model (see Appendix A).

---
[1] Corresponding author. Tel.: +34 942 201635; fax: +34 942 201603.
   E-mail address: sarabiaj@unican.es (J.M. Sarabia).

## 2  Data and Methods

We considered two datasets: the first one corresponds to individual incomes in the year 2012 for United States and, the second one, a truncated dataset which includes only the individual income for taxpayers in the period 2011-12 for United Kingdom. Tables B1 and B2, in Appendix B, show respectively the datasets considered, which come from the U.S. Bureau of Census (2014a, b) and from the U.K. HM Revenue & Customs (2014).

We fitted one-parameter exponential model to these datasets by the method of nonlinear least squares. We also tested the goodness-of-fit of that model by a Kolmogorov–Smirnov test method based on bootstrap resampling (for more details, see Appendix A). Finally, we validated it graphically by comparing the observed data with the theoretical cumulative distribution function (cdf) of the model.

## 3  Results and Discussion

One-parameter exponential distribution can be ruled out as a fit to the empirical individual income data for truncated samples. Table 1 presents the results obtained: the parameter estimates, the empirical *KS* statistics and the bootstrap *p-values* for the U.S. and the U.K datasets. It can be seen that the one parameter model cannot be rejected in the case of the U.S. dataset $(p = 0.9802 > 0.05)$, but it is rejected for the U.K. dataset $(p = 0.0000 \leq 0.05)$. Figure 1 shows graphically the adequacy of the one-parameter exponential model to the U.S. dataset and its poor fit in the case of the U.K.

In contrast, the two-parameter exponential distribution cannot be ruled out as a fit to empirical individual income data, for both datasets considered. Table 2 shows the results for the two-parameter model: the parameter estimates, the empirical *KS* statistics and the bootstrap p-values for the U.S. and the U.K datasets. It can be seen that *p-values* are very close to 1 in both models, supporting the two-parameter model in both cases. Figure 2 confirms graphically the adequacy of the two-parameter exponential model to both datasets. The explanation is that, while the scale parameter $\sigma$ let us to model different countries with different sizes, the location parameter $\theta$ let us to solve the problem of truncated data – the U.K. dataset includes taxpayers only.

## 4  Conclusions

In this paper, we confirm the exponential behavior of the individual income distribution in U.S (2012) and U.K. (2011-12), but considering a two-parameter model instead of one-parameter model considered by Cho (2014); Dragulescu and Yakovenko (2001a, b).

**Table 1:** Parameter estimates, empirical *KS* statistics and bootstrap p-values for one-parameter exponential model of the U.S. (year 2012) and U.K (period 2011-12) datasets. Values of $p \leq 0.05$ indicate that exponential model can be rejected with the 0.05 level of significance.

| Dataset | $\hat{\sigma}$ | KS | $p-value$ | Support for 1-parameter exponential model |
|---|---|---|---|---|
| U.S. (2012) | 38065.8 | 0.0463115 | 0.9802 | OK |
| U.K. (2011-12) | 30678.0 | 0.2129867 | 0.0000 | None |

**Table 2:** Parameter estimates, empirical KS statistics and bootstrap p-values for two-parameter exponential model, of the U.S. (year 2012) and U.K (period 2011-12) datasets. Values of $p \leq 0.05$ indicate that exponential model can be rejected with the 0.05 level of significance.

| Dataset | $\hat{\sigma}$ | $\hat{\theta}$ | KS | $p-value$ | Support for 2-parameter exponential model |
|---|---|---|---|---|---|
| U.S. (2012) | 36059.8 | 1854.97 | 0.0400717 | 0.9887 | OK |
| U.K. (2011-12) | 17506.5 | 8260.56 | 0.0401815 | 0.8311 | OK |

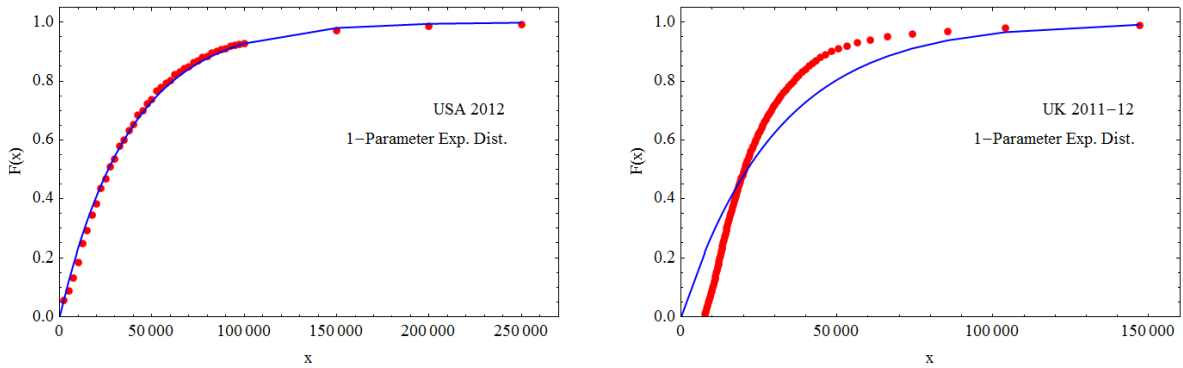

**Fig. 1.** Plot of the cumulative distribution function (cdf) of the 1-parameter exponential distribution (solid lines) and the observed data. Left: U.S. (2012). Right: U.K. (2011-12).

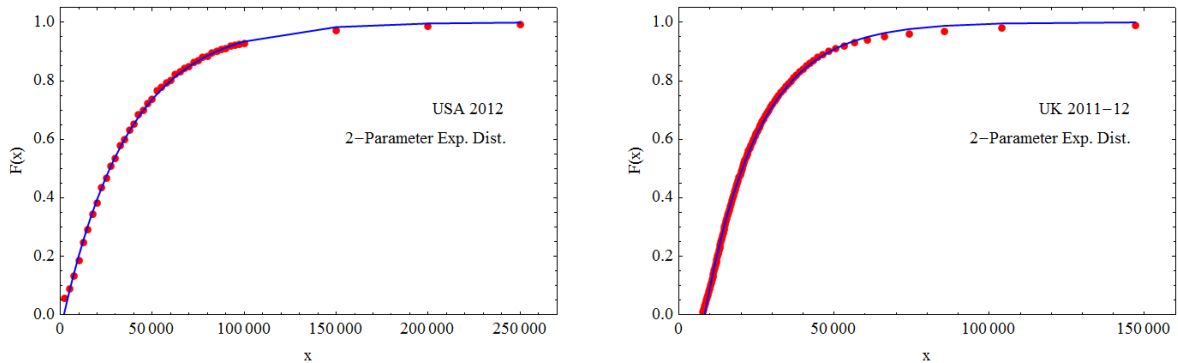

**Fig. 2.** Plot of the cumulative distribution function (cdf) of the 2-parameter exponential distribution (solid lines) and the observed data. Left: U.S. (2012). Right: U.K. (2011-12).

**Acknowledgements:**

The authors thank to Ministerio de Economía y Competitividad (project ECO2010-15455), Ministerio de Educación (FPU AP-2010-4907) and University of Cantabria (Proyectos Puente 2014) for partial support of this work.

**Appendix A: Methods**

- Exponential distribution (Sarabia, 2008).

The classical one-parameter exponential distribution is defined, in terms of the cdf, as follows:
$$F(x) = \Pr(X \leq x) = 1 - \exp(-x/\sigma), \ x > 0, \ \sigma > 0$$

Two-parameter exponential distribution is defined, in term of the cumulative distribution function, as follows:
$$F(x) = \Pr(X \leq x) = 1 - \exp\{-(x-\theta)/\sigma\}, \ x > \theta > 0, \ \sigma > 0.$$

If $\theta$ is the truncation income parameter, the corresponding model is,
$$\{X | X > \theta\} . \tag{A1}$$

The survival function of (A1) is,
$$\Pr(X > x | X > \theta) = \frac{\Pr(X > x)}{\Pr(X > \theta)}, \text{ if } x > \theta.$$

According to the Dragulescu-Yakovenko (2000) model, X is distributed as an exponential distribution and then we have,
$$\Pr(X > x | X > \theta) = \frac{\exp(-x/\sigma)}{\exp(-\theta/\sigma)} = \exp\{-(x-\theta)/\sigma\}, \text{ if } x > \theta.$$

The Lorenz curve corresponding to the truncated model is
$$L(p; \sigma, \theta) = p + \left(1 + \frac{\theta}{\sigma}\right)^{-1} (1-p) \log(1-p), \quad 0 \leq p \leq 1,$$

and the Gini index is
$$G(\theta, \sigma) = 2 \int_0^1 \left[ p - L(p; \theta, \sigma) \right] dp = \frac{\sigma}{2(\sigma + \theta)}.$$

- Nonlinear least squares fit of the exponential distribution:

We fitted the exponential distribution, with one parameter and with two parameters, by solving respectively
$$\min_{\sigma} f(\sigma) = \sum_{i=1}^{n} (F_n(x_i) - F(x_i))^2 \ ;$$

$$\min_{\theta,\sigma} f(\theta,\sigma) = \sum_{i=1}^{n}\left(F_n(x_i) - F(x_i)\right)^2,$$

where $F_n(x_i)$ is the empirical cumulative distribution function (see tables B1,B2), $F(x_i)$ is the theoretical cumulative distribution function mentioned before and *n* is the sample size.

Nonlinear least squares estimates of *σ* (and *θ*) were computed by numerical methods, using the Mathematica (Wolfram Research, Inc., 2010) software function "FindMaximum"; and taking as the initial values: $\hat{\sigma}_0 = \frac{1}{n}\sum_{i=1}^{n}\left(x_i - \hat{\theta}_0\right)$, where $\hat{\theta}_0 = 0$ for the one-parameter model, and $\hat{\theta}_0 = x_{\min}$ (the sample minimum) in the case of two-parameter distribution, which are the maximum likelihood estimators.

- Kolmogorov–Smirnov test method based on bootstrap resampling:

The null hypothesis to test is $H_0$: *the data follow the one-parameter (or two-parameter) exponential model*.

The goodness-of-fit statistic used is the Kolmogorov-Smirnov (*KS*) statistic, given respectively by

$$D_n = \sup\left|F_n(x_i) - F(x_i; \hat{\sigma})\right|;$$
$$D_n = \sup\left|F_n(x_i) - F(x_i; \hat{\sigma}, \hat{\theta})\right|; \quad i = 1, 2, \ldots, n,$$

The procedure is as follows (Prieto et al, 2014; Clauset et al., 2009): calculate the empirical *KS* statistic for the observed data; generate, by simulation, enough synthetic data sets (in this study, we generated 10000 data sets), with the same sample size *n* as the observed data; fit each synthetic data set by nonlinear least squares method and obtain its theoretical cdf; calculate the *KS* statistic for each synthetic data set – with its own theoretical cdf; calculate the *p-value* as the fraction of synthetic data sets with a *KS* statistic greater than the empirical *KS* statistic; and exponential model can be ruled out if $p-value \leq 0.05$.

## Appendix B: Datasets

**Table B1 -** Total income, expressed in US dollar ($), and the corresponding empirical cumulative distribution function, for people with income in U.S, in the year 2012, published by U.S. Bureau of Census.

| x | 2500 | 4999 | 7499 | 9999 | 12499 | 14999 | 17499 | 19999 | 22499 | 24999 | 27499 | 29999 | 32499 |
|---|---|---|---|---|---|---|---|---|---|---|---|---|---|
| F(x) | 0.0578 | 0.0903 | 0.1325 | 0.1858 | 0.2466 | 0.2903 | 0.3435 | 0.3829 | 0.4340 | 0.4659 | 0.5088 | 0.5340 | 0.5787 |

| x | 34999 | 37499 | 39999 | 42499 | 44999 | 47499 | 49999 | 52499 | 54999 | 57499 | 59999 | 62499 | 64999 |
|---|---|---|---|---|---|---|---|---|---|---|---|---|---|
| F(x) | 0.5980 | 0.6327 | 0.6509 | 0.6854 | 0.6990 | 0.7237 | 0.7378 | 0.7673 | 0.7773 | 0.7934 | 0.8015 | 0.8225 | 0.8298 |

| x | 67499 | 69999 | 72499 | 74999 | 77499 | 79999 | 82499 | 84999 | 87499 | 89999 | 92499 | 94999 | 97499 |
|---|---|---|---|---|---|---|---|---|---|---|---|---|---|
| F(x) | 0.8434 | 0.8498 | 0.8645 | 0.8697 | 0.8804 | 0.8849 | 0.8959 | 0.8999 | 0.9063 | 0.9098 | 0.9178 | 0.9210 | 0.9248 |

| x | 99999 | 149999 | 199999 | 249999 |
|---|---|---|---|---|
| F(x) | 0.9275 | 0.9722 | 0.9863 | 0.9918 |

**Table B2** – Total individual income before tax, expressed in pound sterling (£), and the corresponding empirical cumulative distribution function, for taxpayers only, in U.K, in the period 2011-12, published by U.K. HM Revenue & Customs.

| x | 7740 | 8000 | 8280 | 8560 | 8840 | 9150 | 9450 | 9740 | 10000 | 10200 | 10400 | 10700 | 10900 |
|---|---|---|---|---|---|---|---|---|---|---|---|---|---|
| F(x) | 0.01 | 0.02 | 0.03 | 0.04 | 0.05 | 0.06 | 0.07 | 0.08 | 0.09 | 0.10 | 0.11 | 0.12 | 0.13 |

| x | 11100 | 11300 | 11500 | 11700 | 12000 | 12200 | 12400 | 12600 | 12900 | 13100 | 13300 | 13500 | 13800 |
|---|---|---|---|---|---|---|---|---|---|---|---|---|---|
| F(x) | 0.14 | 0.15 | 0.16 | 0.17 | 0.18 | 0.19 | 0.20 | 0.21 | 0.22 | 0.23 | 0.24 | 0.25 | 0.26 |

| x | 14000 | 14300 | 14500 | 14700 | 15000 | 15200 | 15500 | 15800 | 16000 | 16300 | 16300 | 16800 | 17100 |
|---|---|---|---|---|---|---|---|---|---|---|---|---|---|
| F(x) | 0.27 | 0.28 | 0.29 | 0.30 | 0.31 | 0.32 | 0.33 | 0.34 | 0.35 | 0.36 | 0.37 | 0.38 | 0.39 |

| x | 17400 | 17600 | 17900 | 18200 | 18500 | 18800 | 19100 | 19400 | 19700 | 20000 | 20300 | 20700 | 21000 |
|---|---|---|---|---|---|---|---|---|---|---|---|---|---|
| F(x) | 0.40 | 0.41 | 0.42 | 0.43 | 0.44 | 0.45 | 0.46 | 0.47 | 0.48 | 0.49 | 0.50 | 0.51 | 0.52 |

| x | 21300 | 21700 | 22100 | 22400 | 22800 | 23200 | 23600 | 24000 | 24400 | 24900 | 25300 | 25800 | 26300 |
|---|---|---|---|---|---|---|---|---|---|---|---|---|---|
| F(x) | 0.53 | 0.54 | 0.55 | 0.56 | 0.57 | 0.58 | 0.59 | 0.60 | 0.61 | 0.62 | 0.63 | 0.64 | 0.65 |

| x | 26800 | 27300 | 27800 | 28400 | 29000 | 29500 | 30100 | 30800 | 31400 | 32100 | 32800 | 33600 | 34400 |
|---|---|---|---|---|---|---|---|---|---|---|---|---|---|
| F(x) | 0.66 | 0.67 | 0.68 | 0.69 | 0.70 | 0.71 | 0.72 | 0.73 | 0.74 | 0.75 | 0.76 | 0.77 | 0.78 |

| x | 35200 | 36000 | 36900 | 37900 | 39000 | 40000 | 41100 | 42200 | 43400 | 44800 | 46400 | 48300 | 50500 |
|---|---|---|---|---|---|---|---|---|---|---|---|---|---|
| F(x) | 0.79 | 0.80 | 0.81 | 0.82 | 0.83 | 0.84 | 0.85 | 0.86 | 0.87 | 0.88 | 0.89 | 0.90 | 0.91 |

| x | 53200 | 56500 | 60700 | 66200 | 74100 | 85500 | 104000 | 147000 |
|---|---|---|---|---|---|---|---|---|
| F(x) | 0.92 | 0.93 | 0.94 | 0.95 | 0.96 | 0.97 | 0.98 | 0.99 |

# References

Cho, A. *Physicists say it's simple,* Science, **344** (6186), 828 (2014).

Clauset, A, Shalizi CR, Newman MEJ. *Power-law distributions in empirical data*. SIAM Rev 2009;51(4):661–703.

Dragulescu, A., Yakovenko, V.M., Statistical mechanics of money, Eur. Phys. J. B. **17**, 723-729 (2000).

Dragulescu, A., Yakovenko, V.M., *Exponential and power-law probability distributions of wealth and income in the United Kingdom and the United States*, Physica A **299**, 213-221 (2001a).

Dragulescu, A., Yakovenko, V.M., Evidence for the exponential distribution in income in the USA, Eur. Phys. J. B. **20**, 585-589 (2001b).

Prieto, F., Sarabia, J.M., Sáez, A.J., *Modelling major failures in power grids in the whole range*. Int. J. Elec. Power Energy Syst. **54**, 10–16 (2014).

Sarabia, J.M. (2008). "Parametric Lorenz Curves: Models and Applications". In: Modeling Income Distributions and Lorenz Curves. Series: Economic Studies in Inequality, Social Exclusion and Well-Being 4, Chotikapanich, D. (Ed.), pp. 167-190, Springer-Verlag.

U.K. HM Revenue & Customs, https://www.gov.uk/government/publications/percentile-points-from-1-to-99-for-total-income-before-and-after-tax . Accessed 15 June 2014.

U.S. Bureau of Census, Accessed 15 June 2014.
http://www.census.gov/hhes/www/cpstables/032013/perinc/pinc01_000.htm .

U.S. Bureau of Census, Accessed 15 June 2014.
http://www.census.gov/hhes/www/cpstables/032013/perinc/pinc11_000.htm .

Wolfram Research, Inc., 2010. Mathematica. Version 8.0. Wolfram Research, Inc.,Champaign, IL.
7